\documentclass[11pt,twoside]{article}

\usepackage{asp2006}
\usepackage{epsf}
\usepackage{epsfig}
\usepackage{lscape}
\def\kms{km~s$^{-1}$}
\def\sun{$_{\rm \odot}$}
\begin{document}
\title{Metal Enriched Outflows and Dusty starburst in the type
       II quasar Q 1321+058}   
\author{Ting-Gui Wang$^1$, Hongyan Zhou$^1$, Honglin Lu$^1$, Weimin Yuan$^2$,
        Hongguang Shan$^2$ \& Xiaobo Dong$^2$}   
\affil{1.Center for Astrophysics, University of Science and
Technology of China, Hefei, 230026, China}    
\affil{2.National Astronomical Observatories/Yunnan
Observatory, CAS, Kunming, Yunnan, P.O.
BOX 110, P.R.China}

\begin{abstract}
We present an analysis of the
emission line property and the broad band spectral
energy distribution of the ultra-luminous infrared \textit{Type II}
quasar Q132+058. 
The optical and ultraviolet emission lines show four distinct
components: a LINER-like component at the systematic velocity, a
heavily reddened HII-like component blueshifted by about 400\kms,
and two broad components blueshifted by about 400\kms and 1900\kms,
respectively. The emission line ratios suggest that broad components
are produced in dense and $\alpha$-enriched outflows with a
metalicity of 3-5Z\sun~and a density of $n_H\sim 10^7~cm^{-3}$. The
optical--UV continuum is  dominated by stellar light and can be
modeled with a young ($<1$~Myr) plus an intermediate age
(0.5-0.8Gyr) stellar populations. The near to mid-infrared light is
dominated by hot and warm dust heated by the hidden quasar. We
derive a star formation rate (SFR) of 300-500 M\sun~yr$^{-1}$ from the
UV spectrum and far-infrared luminosity, which is two orders of
magnitude larger than that indicated by reddening uncorrected [OII]
luminosity.
\end{abstract}

\vspace{-1.2cm}
\section{Introduction}   

Mounting evidence has been found 
in the past decade that quasar activity is
triggered through merging of two gas-rich massive galaxies.
Merging also leads to starburst in both the nucleus and
the extended region. As such, most of these systems show up as 
luminous or ultra-luminous infrared galaxies (ULIRGs, $L_{ir}>10^{12}L$\sun).
The connection between ULIRGs and quasars has been widely discussed,
and there is evidence for an evolution sequence between the two (e.g., Sanders et al. 1996).

We present here only a brief account of the
results from an analysis of the optical/ultraviolet spectra and
the broad band properties of the ULIRG Q 1321+058. 
Detailed analysis and discussion will be presented elsewhere
(Wang et al. 2007, in preparation). The object was
classified as a quasar based on a low resolution spectrum taken 
in an attempt of optical identification of 
HEAO-A2 X-ray sources (Remillard et al. 1993).
However, it was not detected by XMM Newton (Bianchi et al. 2005), which led
those authors to propose that it is a Compton-thick type II object. 
Its HST images show an elongated tidal tail and a bright nucleus, indicating
a merger at a later stage.

\section{Data analysis and results}

Its SDSS spectrum and HST FOS spectrum were extracted from the archives
and were corrected for
the Galactic reddening. The emission
line profile of [OIII] is complex and is
clearly resolved into multiple components,
as shown by Lipari et al. (2003). 
The similar line profiles of 
H$\beta$ and H$\alpha$ to that of the [OIII] line
rule out that it is a Type I AGN.
The optical continuum is dominated by stellar light
with noticeable absorption in high order Balmer lines.
The broad UV and optical continuum can be fairly well fitted by a
model composed of two reddened stellar populations: an intermediate-aged
stellar population of 0.5-0.8 Gyr and a young stellar population
with an age $<$ 1 Myr using the stellar templates STARBURST99 (Leither et 
al. 1999;) in UV and the single stellar population
spectra in GALAXEV (Bruzual \& Charlot 2003) in optical. For a Salpeter
initial mass function that extends to 0.08M\sun, the stellar mass of
the two populations are 4.5$\times10^8$ and
5.8$\times10^{10}$ M\sun, respectively.

The emission line profiles show four distinct components. The
CIV$\lambda\lambda$1548, 1550 and NIII]$\lambda$1750 show
single-peak profiles blueshifted by $\sim$ 2,000 \kms
relative to the systematic redshift determined from the stellar
absorption lines, and can be well fitted with a single Gaussian
(see Fig \ref{fig3}). [OII]$\lambda\lambda$ 3726, 3729,
[OI]$\lambda$6302 and [SII]$\lambda\lambda$6713, 6731 display a
single-peaked profile at the systematic velocity. The [NII]
line may also be single peaked but badly blended with
the H$\alpha$ line. The [OIII]$\lambda$4959, 5007, H$\beta$,
[NeIII]$\lambda$3896 lines show two more components in addition to
the above two: a narrow component and a second broad component
blue-shifted by $\sim$400 \kms relative to the systematic velocity.
The less blueshifted broad-component is also found in  AlIII,
CIII], SiIII], H$\alpha$. We refer the four components to as: C1
-- the narrow component at the systematic velocity; C2--the narrow
blueshifted component; C3--the broad component with a smaller
blueshift; C4--the broad component with a large bueshift.

\begin{figure}[!ht]
\vspace{-5.5cm}
\includegraphics[scale=0.74]{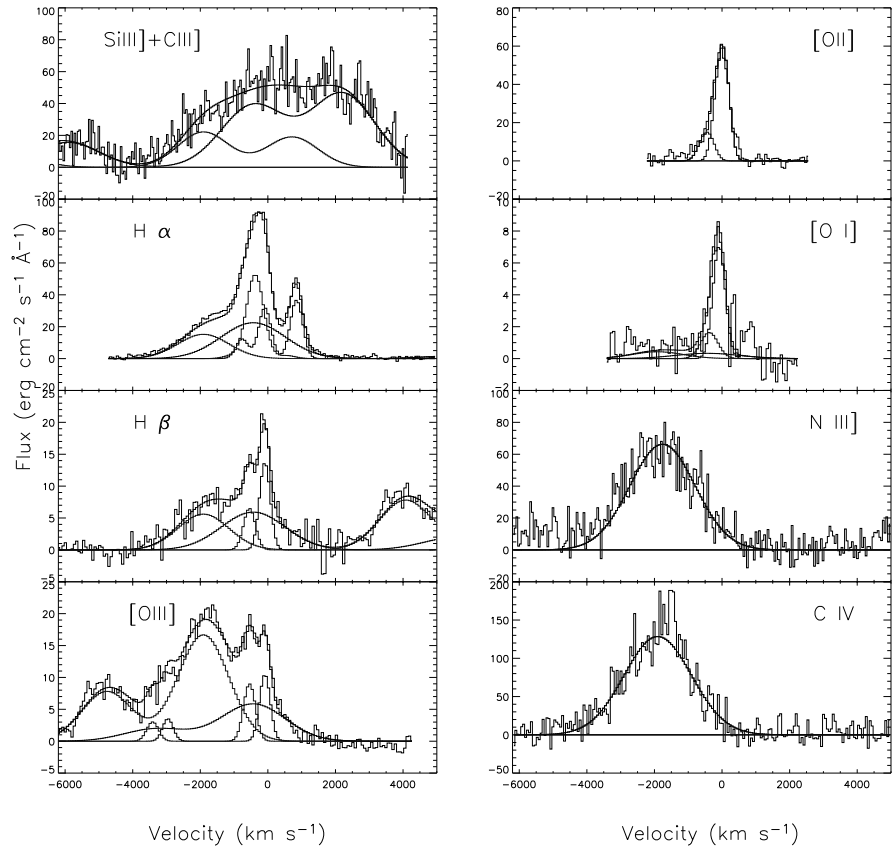}
\vspace{-0.5cm}
\caption{Emission line profiles and their best fitted four component
models (refer to text for detail). \label{fig1}}
\vspace{-0.3cm}
\end{figure}

We fit the emission line profiles with 4 gaussian components (Fig
\ref{fig1}). The C1 component locates in the LINER region on the BPT
diagram, while C2 is in the HII regime and is highly reddened.
It is noted that both SiIII]/CIII] $\simeq$1 and
NIII]/CIV$\simeq$ 0.56 are among the highest known values for
quasars.

\section{Discussions}

\begin{figure}[!ht]
\vspace{-0.2cm}
\includegraphics[scale=0.60]{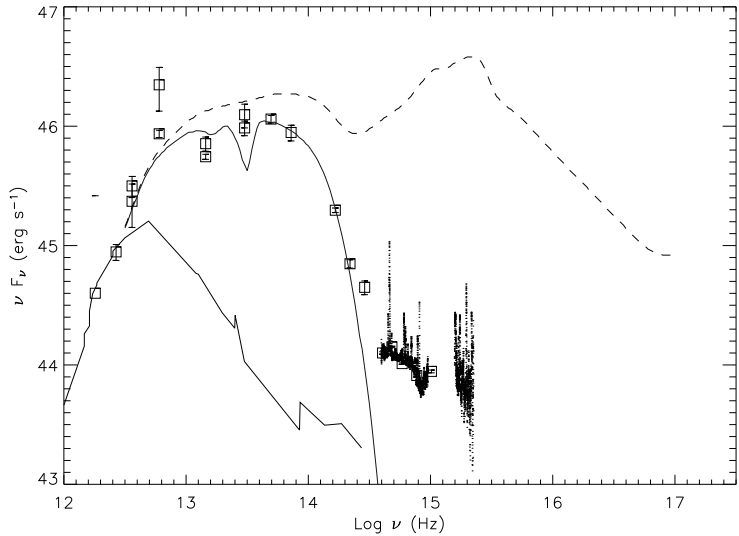}\vspace{-0.2cm}
\caption{Spectral energy distribution from far-infrared to the ultraviolet
for Q1321+058. The infrared SED of NGC 6240 is shown in solid line, the
template of infrared-luminous quasars in Richards et al. (2006) in dashed- 
curve, while the quasar template reddened by E(B-V)=4.5 in thin-curve.
\label{fig2}}
\end{figure}

The broad band spectral energy distribution (SED) of Q
1321+058 as shown in Fig \ref{fig2} is consistent with that of
a Type II quasar: the steep rise from near to mid-infrared and no apparent
absorption or emission features around 10 $\mu$m suggest that the
mid-infrared emission is dominated by an active nucleus. The
integrated infrared luminosity is around 2$\times10^{46}$
erg~s$^{-1}$. The far-infrared luminosity is similar to that of NGC
6240. If most far-infrared emission is attributed to star formation
(Schweitzer et al. 2006), the star formation rate (SFR)
will be about 300M\sun~yr$^{-1}$, comparable to that
inferred from the young stellar population. However,
the luminosity of the narrow
[OII] component gives an upper limit of SFR 1.5 M\sun~yr$^{-1}$,
which is two orders of
magnitudes lower than the SFR in the galaxy.
This can be explained if the star-formation is obscured, 
which is also suggested by the reddening of the the narrow components
based on their Balmer decrements.
Applying extinction correction using the Balmer decrement 
brings the two \textit{SFRs} to good agreement.

We used version C06.02c of Cloudy, last described by (Ferland 1998),
to explore the physical parameters of the emission
regions by  re-producing the observed line ratios for the
broad components. For the solar abundances or scaled solar
abundances, a large SiIII]/CIII] value can be reproduced only for
high gas density ($n_H>10^{9.5}$cm $^{-3}$), while a much lower
density ($n_H\sim 10^7$ cm$^{-3}$) is required to explain the
observed the [OIII]$\lambda$4363/[OIII]$\lambda$5007 ratio and the
[NeIII]$\lambda$3896/[OIII]$\lambda$5007 ratio. 
Multiple phase
medium has the difficulty that the ionization parameters
for the two phases would be largely different 
under the reasonable assumption that they are exposed to the
same ionizing continuum.
A self-consistent solution can be found if the
metal abundance is 5 times the solar value for a constant density 
model with a metal enriched
scheme outlined by Hamann and Ferland (1993). The density of the
emission line gas is around $10^7$~cm$^{-3}$, and the ionization
parameters around U=10$^{-2}$ for C3 and 10$^{-1.5}$ for C4. This
places the emission line gas at a distance from the AGN of $\sim$
500 pc for the C3 and $\sim$ 200 pc for the C4 emission region,
provided that the AGN has the broad band SED similar to the 
average of blue quasars in Richards et al. (2006). These
distances are likely outside the dusty torus, which explains their
low extinctions.

The projected velocity of the blue component
reaches $\sim$4,000 km~s$^{-1}$ at the blue wing for the C4
component. Q 1321+058 would appear as a BAL QSO
provided that the line of sight intersected the outflow. We
estimate from the broad H$\alpha$ luminosity a minimum mass 
to be
$\sim$600 M\sun for the C3 and $\sim$300 M\sun for the C4 components, 
assuming the
emission line gas is HII dominated and Case-B recombination. The
mass loss rate in this cold component is not important if most of
the gas is visible on a dynamic time scale. However, the filling factor
is extremely small ($\sim 10^{-11}$), indicating a possible
two-phase medium in which the hot phase gas dominates the cold one.

\acknowledgements This work is supported by Chinese Natural Science
Foundation through CNSF-10233030, CNSF-10573015 and CNSF-10473013.

\end{document}